\documentclass{svproc}
\usepackage{url}

\usepackage{amsmath,latexsym,amssymb,graphicx,dsfont,amsfonts}

\DeclareMathOperator*{\argmin}{arg\,min}
\begin{document}
\mainmatter              
\title{Fast Probabilistic Consensus with Weighted Votes}
\titlerunning{Fast Probabilistic Consensus with Weighted Votes}  
%
\author{Sebastian M\"uller\inst{*,1} \and Andreas Penzkofer\inst{*,2} \and Bartosz Ku\'{s}mierz\inst{2} \and Darcy Camargo\inst{2,3} \and William J.~Buchanan\inst{4}}
%
\authorrunning{M\"uller, Penzkofer, Ku\'{s}mierz,  Camargo, Buchanan} 
%
%
\institute{Aix Marseille Universit\'e, CNRS, Centrale Marseille, I2M - UMR 7373, 13453 Marseille, France,\\
\email{sebastian.muller@univ-amu.fr}
\and
IOTA Foundation, 10405 Berlin, Germany\\
\email{ \{andreas.penzkofer, bartosz.kusmierz, darcy.camargo\}@iota.org}
\and
Department of Mathematics, Weizmann Institute, POB 26, Rehovot 7610001, Israel
\and Blockpass ID Lab, Edinburgh Napier University, Edinburgh, UK
\email{ b.buchanan@napier.ac.uk}}

\maketitle              

\begin{abstract}
The fast probabilistic consensus (FPC) is a voting consensus protocol that is robust and efficient in Byzantine infrastructure.  We propose an adaption of the FPC to a setting where the voting power is proportional to the nodes reputations. We model the reputation using a Zipf law and show using simulations that the performance of the protocol in Byzantine infrastructure increases with the Zipf exponent. Moreover, we propose several improvements of the FPC that decrease the failure rates significantly and  allow the protocol to withstand adversaries with higher weight. We distinguish between cautious and berserk strategies of the adversaries and propose an efficient method to detect the more harmful berserk strategies. Our study refers at several points to a specific implementation of the IOTA protocol, but the principal results hold for general implementations of reputation models.\keywords{Distributed systems, consensus protocols, fairness, Sybil attack, Byzantine infrastructures, simulation studies }
\end{abstract}

\renewcommand{\thefootnote}{\fnsymbol{footnote}}
\setcounter{footnote}{0}
\footnotetext{$^{*}$These authors contributed equally.}
\renewcommand*{\thefootnote}{\arabic{footnote}}
\setcounter{footnote}{0}

\section{Introduction}
Distributed consensus algorithms allow networked systems to agree on a required state or opinion in situations where centralized decision making is difficult or even impossible. As distributed computing is inherently unreliable, it is necessary to reach consensus in faulty or Byzantine infrastructure. The importance of this problem stems from its omnipresence and fault tolerance is one of the most fundamental aspects of distributed computing, e.g.,~\cite{BaDaMa:93}.

This article focuses on a consensus protocol that falls into the class of binary majority voting consensus protocols. The basic idea is that nodes query other nodes about their current opinion, and adjust their own opinion over the course of several rounds based on the proportion of other opinions they have observed. The functional principle of this protocol, already observed  by the Marquis de Condorcet in 1785 \cite{marquis}, relies on the law of large numbers; suppose there is a large population of voters, and each one independently votes "correctly" with probability $p > 1/2$. Then as the population size grows, the probability that the outcome of a majority vote is "correct" converges to one.

While voting consensus protocols have their limitations, they have been successfully applied not only in decision making but also in a wide range of engineering and economical applications
,  and lead to the emerging science of sociophysics~\cite{castellano2009}. 
 
We continue the works of \cite{fpc} and \cite{fpcsim} and propose several adaptions, Section \ref{sec:improvements}, of the fast probabilistic consensus protocol (FPC) that decreases the failure rate of at least one order of magnitude, e.g.,~see Fig. \ref{fig:N}. The main contribution is the adaption of the protocol to a setting allowing defense against Sybil attacks.   

In FPC nodes need to be able to query a sufficiently large proportion of the network directly, which requires that nodes have global identities (node IDs) with which they can be addressed. In a decentralized and permissionless setting  a malicious actor may gain a disproportionately large influence on the voting by creating a large number of pseudonymous identities.
In the blockchain environment, mechanisms such as proof-of-work and (delegated) proof-of-stake can act as a Sybil mitigation mechanism in the sense that the voting power is proportional to the work invested or the value staked \cite{blockchain_sybil}.

For the IOTA protocol \cite{coordicide} introduces \textit{mana} as a  Sybil defense, where mana is delegated to nodes and proportional to the active amount of IOTA in the network. While in the remainder of the paper we will always refer to mana, the protocol can be implemented using any \emph{good} or \emph{resources} that can be verified via resource testing or recurring costs and fee, e.g.,~\cite{Neil_asurvey}. 
In Section \ref{sec:fairness} we propose a weighted voting consensus protocol that is fair in the sense that the voting power is proportional to the nodes' reputation. 

\begin{center}
\begin{figure}[ht]
\centering
    \includegraphics[width=0.8\textwidth,trim={0 0 0 0cm},clip]{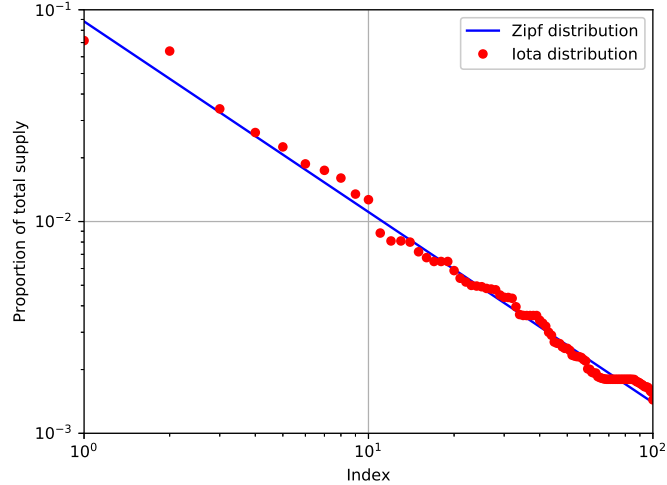}
    \caption{Distribution of relative IOTA value on the top 100 addresses with a fitted Zipf distribution with $s=0.9$.}
    \label{fig:IOTAzipf}
\end{figure}
\end{center}

In general, values in (crypto-)currency systems are not distributed equally; \cite{btcdistribution} investigates the heterogeneous distribution of the wealth across Bitcoin addresses and finds that it follows certain power laws.  Power laws satisfy a universality phenomenon; they appear in numerous different fields of applications and have, in particular, also been utilised to model wealth in economic models \cite{wealth_pareto}. In this paper we consider a Zipf law to model the proportional wealth of nodes in the IOTA network: the $n$th largest value $y(n)$ satisfies
\begin{equation}\label{eq:zipf}
    y(n)=C n^{-s},
\end{equation}
where $C^{-1}=\sum^N_{n=1} n^{-s}$, $N$ is the number of nodes, and $s$ is the Zipf parameter. Fig. \ref{fig:IOTAzipf} shows the distribution of IOTA for the top 100 richest addresses\footnote{\url{https://thetangle.org}\label{footnoteIOTA}} together with a fitted Zipf distribution. Since (\ref{eq:zipf}) only depends on two parameters, $s$ and $N$, this provides a convenient model to investigate the performance of FPC in a wide range of network situations. For instance, networks where nodes are equal may be modelled by choosing $s=0$, while more centralized networks can be considered for $s>1$. We refer to Section \ref{sec:Zipf} for more details on the Zipf law.

\subsection*{Outline}
The rest of the paper organizes as follows. After giving an introduction to the original version of FPC  in Section \ref{sec:vanillaFPC}, we summarize results on the fairness of this protocol in Section \ref{sec:fairness}.  In Section \ref{sec:Zipf} we propose modelling of the weight distribution using a Zipf law, we highlight the skewness of this distribution in Section \ref{sec:skewness}, and in  Section \ref{sec:complexity} we discuss how the properties of the Zipf law influence the message complexity of the protocol.

After defining the threat model in Section \ref{sec:threat} we propose several improvements of the Vanilla FPC in Section \ref{sec:improvements}. In Section \ref{sec:berserk}, we outline a protection mechanism against the most severe attack strategies.  The quorum size is an important parameter of FPC that dominates its performance; we give in Section \ref{sec:k} a heuristic to choose a quorum size for a given security level.

Section \ref{sec:simresults} presents simulation results that show the performance of the protocol in Byzantine infrastructure for different degrees of centralization of the weights. We conclude in Section \ref{sec:discussion} with a discussion.

\section{Vanilla FPC}\label{sec:vanillaFPC}
We present here only the key elements of the proposed
protocol and refer the interested reader to~\cite{fpc} and \cite{fpcsim} for more details. In order to define FPC we have to  introduce some notation.
We assume the network to have $N$ nodes indexed by $1,2,\ldots, N$ and that every node is able to query any other nodes.\footnote{This assumption is only made for sake of a better presentation; a node does not need to know every other node in the network. While the theoretical results in \cite{fpc} are proven under this assumption, simulation studies \cite{fpcsim} indicate that it is sufficient if every node knows about half of the other nodes. Moreover, it seems to be a reasonable assumption that large mana nodes are known to every participant in the network.}  Every node $i$ has an opinion or state. We note $s_{i}(t)$ for the opinion of the node $i$ at time $t$. Opinions take values in $\{0,1\}$. Every node $i$ has an initial opinion $s_{i}(0)$.

At each (discrete) time step each node chooses $k$ random nodes $C_{i}=C_i(t)$, queries their opinions and calculates 
\begin{equation*}
\eta_{i}(t+1)=\frac1{k_{i}(t)} \sum_{j\in C_{i}} s_{j}(t),
\end{equation*}
where $k_{i}(t)\leq k$ is the number of replies received by node $i$ at time $t$ and $s_j(t)=0$ if the reply from $j$ is not received in due time. 
Note that the neighbors $C_{i}$ of a node $i$ are chosen using sampling with replacement and hence repetitions are possible.

As in \cite{fpcsim} we consider a basic version of the FPC introduced in \cite{fpc} in choosing some parameters by default. Specifically, we remove the cooling phase of FPC and the randomness of the initial threshold $\tau$. Let $U_{t}$, $t=1, 2,\ldots$ be i.i.d.~random variables with law $\mathrm{Unif}( [\beta, 1-\beta])$ for some parameter $\beta \in [0,1/2]$. The update rules for the opinion of a node $i$ is then given by
\begin{equation*}
s_{i}(1)=\left\{ \begin{array}{ll}
1, \mbox{ if } \eta_{i}(1) \geq \tau, \\
0, \mbox{ otherwise,}
\end{array}\right. 
\end{equation*}
and for $t\geq 1$:
\begin{equation*}
s_{i}(t+1)=\left\{ \begin{array}{ll}
1, \mbox{ if } \eta_{i}(t+1) > U_{t}, \\
0, \mbox{ if } \eta_{i}(t+1) < U_{t}, \\
s_{i}(t), \mbox{ otherwise.}
\end{array}\right. 
\end{equation*}
Note that if $\beta=0.5$, FPC reduces to a standard majority consensus. The above sequence of random variables $U_t$ are the same for all nodes; we refer to \cite{fpcsim} for a more detailed discussion on the use of decentralized random number generators.

We introduce a local termination rule to reduce the communication complexity of the protocols. Every node keeps a counter variable \verb?cnt? that is incremented by $1$ if there is no change in its opinion and that is set to $0$ if there is a change of opinion. Once the counter reaches a certain threshold $\verb?l?$, i.e.,~$\verb?cnt?\geq \verb?l?$,  the node considers the current state as final. The node will therefore no longer send any queries but will still answer incoming queries. In the absence of autonomous termination the algorithm is halted after $\verb?maxIt?$ iterations.

\section{Fairness}\label{sec:fairness}
Introducing mana as a weighting factor may naturally have an influence on the mana distribution and may lead to degenerated cases. In order to avoid this phenomenon we want  to ensure that no node can increase  its importance in splitting up into several nodes, nor can achieve better performance in pooling together with other nodes. 

We consider a network of $N$ nodes whose  mana is described by $\{m_1,..,m_N\}$ with $\sum^N_{i=1} m_i=1$. In the sampling of the queries a  node $j$ is chosen now with probability  
$$
p_j=\frac{f(m_j)}{\sum_{i=1}^N f(m_i)}.
$$ 
Each opinion is weighted by $g_j=g(m_j)$, resulting in the value
$$\eta_i(t+1)= \frac{1}{\sum_{j\in C_i} g_j} \sum_{j\in C_i} g_j s_j(t).
$$
The other parts of the protocol remain unchanged. 

We denote by $y_i$ the number of times a node $i$ is chosen. 
As the sampling is described by a multinomial distribution we can calculate the expected value of a query as
$$
\mathbb{E}\eta(t+1)=\sum_{i=1}^N s_i(t)v_i,
$$
where
$$
v_i=\sum_{\textbf{y}\in \mathbb{N}^N:\sum{y_i}=k}
\frac{k!}{y_1!\cdot\cdot\cdot y_N!}
\frac{y_i g_i }{\sum_{n=1}^N y_n g_n}
\prod^N_{j=1}p_j^{y_j}
$$
is called the voting power of node $i$. The voting power measures the influence of the node $i$.  We would like the voting power to be proportional to the mana. 
\begin{definition}
A voting scheme $(f,g)$ is fair if the voting power is not sensitive to splitting/merging of mana, i.e.,~if a node $i$ splits into nodes $i_1$ and $i_2$ with a mana splitting ratio $x\in (0,1)$, then
\begin{equation}\label{eq:fair}
   v_i(m_i)=v_{i_1}(xm_i)+v_{i_2}((1-x)m_i) 
\end{equation}
\end{definition}
In the case where $g\equiv 1$, i.e.,~the $\eta$ is an unweighted mean,  the existence of a  voting scheme that is fair for all possible choices of $k$ and mana distributions is shown in \cite{Fairness}:
\begin{lemma}\label{lemma1}
For $g\equiv 1$ the voting scheme $(f,g)$ is fair if and only if $f$ is the identity function $f=id$.
\end{lemma}
For this reason we fix from now on $g\equiv 1$ and $f=id$.

\section{Zipf's law and mana distribution}\label{sec:Zipf}
	
One of the most intriguing  phenomenon in probability theory is that of universality;  many seemingly unrelated probability distributions, which may involve large numbers of unknown parameters, can end up converging to a universal law that only depends on few parameters. Probably the most famous  example of this universality phenomenon is the central limit theorem. 

Analogous universality phenomena also show up in empirical distributions, i.e.,~distributions of statistics from a large population of real-world objects. Examples include Benford's law, Zipf's law, and the Pareto distribution\footnote{Interesting to note here that these three distributions are highly compatible with each other.}; we refer to \cite{Zipf} for more details. These laws govern the asymptotic distribution of many statistics which
\begin{enumerate}
    \item take values as positive numbers;
    \item range over many different orders of magnitude;
    \item arise from a complicated combination of largely independent factors; and
    \item have not been artificially rounded, truncated, or otherwise constrained in size.
\end{enumerate}
Out of the three above laws, the Zipf law is the appropriate variant for modelling the mana distribution. The  Zipf law is defined as follows: The $n$th largest value of the statistic $X$ should obey an approximate power law, i.e.,~it should be approximately $C n^{-s}$ for the first few $n=1,2,3,\ldots$ and some parameters $C, s > 0$.

The Zipf law is used in various applications.   For instance,  Zipf's law and the closely related Pareto distribution can be used to mathematically test various models of real-world systems (e.g.,~formation of astronomical objects, accumulation of wealth and population growth of countries). 
An important point is  that Zipf's law does in general not apply on the entire range of $X$, but only on the upper tail region when $X$ is significantly higher than the median; in other words, it is a law for the (upper) outliers of $X$.

The Zipf law tends to break down if one of the hypotheses 1) - 4) is dropped. For instance, if the statistic $X$ concentrates around its mean and does not range  over many orders of magnitude, then the normal distribution tends to be a much better model. If instead the samples of the statistics are highly correlated with each other, then other laws can arise, as for example, the Tracy-Widom law. 

Zipf's law is most easily observed by plotting the data on a log-log graph, with the axes being log(rank order) and log(value). The data conforms to a Zipf law to the extent that the plot is linear and the value of $s$ can be found using linear regression. For instance, Fig. \ref{fig:IOTAzipf} shows the distribution of IOTA for the top 100 richest addresses.

Due to universality phenomemon, the plausibility of hypotheses 1) - 4) above and Fig. \ref{fig:IOTAzipf} we assume a Zipf law for the mana distribution. In Section \ref{sec:discussion} we give more details on the validity of the model.

\section{Skewness of mana distribution}\label{sec:skewness}

For $s>0$ the majority of the nodes would have a mana value less than the average and hence, in the case of an increasing function $f$, 
these nodes would be queried less than in a homogeneous distribution. As a consequence the initial opinion of small mana nodes may become negligible. 

\begin{figure}
\centering
    \includegraphics[width=0.8\textwidth,trim={0 0 0 0cm},clip]{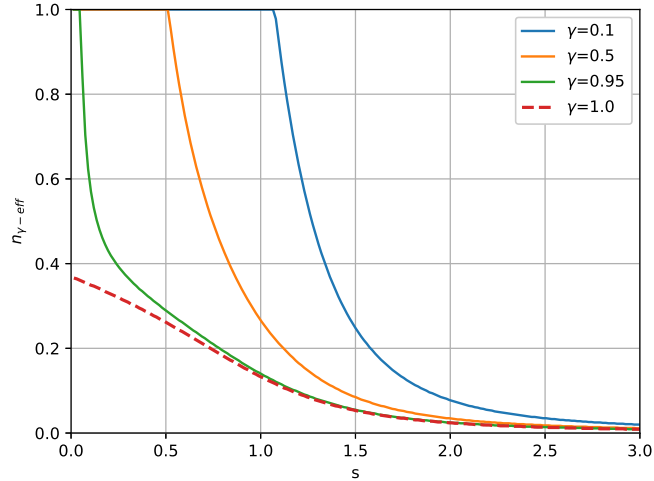}
    \caption{Proportion of effective number of nodes. }
    \label{fig:N_eff}
\end{figure}

We define  the $\gamma$-effective number of nodes $N_{\gamma\text{-eff}}$ as the number of nodes whose proportional mana is more than or equal to $\gamma/N$: 
$$
N_{\gamma\text{-eff}}=\sum_{i=1}^N \mathbf{1}\{m_i \geq \gamma/N\}
$$
where  $\mathbf{1}$ is the standard indicator function.
Fig. \ref{fig:N_eff} shows the relative proportion of effective nodes $n_{\gamma\text{-eff}}=N_{\gamma\text{-eff}}/N$ with $s$. We show the figure for $N=1000$, although the distribution hardly changes when changing $N$. Note that for $\gamma=1$ and $s\rightarrow 0$ a large proportion of the nodes would have less than a proportion $1/N$ of the mana and hence $n_{\gamma\text{-eff}}$ approaches, as $s\to 0$, to a value strictly less than $1$.
Note that for values of $s\gtrapprox1$ the effective number of nodes can be very small. This is also reflected in the distribution of IOTA. The top 100 addresses shown in Fig. \ref{fig:IOTAzipf} own $~60\%$ of the total funds, albeit there are more than 100.000 addresses in total$^\text{\ref{footnoteIOTA}}$.

\section{Message complexity}\label{sec:complexity}
 Let us start with the following back-of-the-envelope calculation. Denote by $h(N)$ the mana rank of a given node. At every round this node is queried on average 
\begin{equation}
    N\cdot  \frac{h(N)^{-s}}{\sum_{n=1}^N n^{-s}}
\end{equation} times.
Now, if $s< 1$ this becomes asymptotically $\Theta(N^s h(N)^{-s})$, if $s=1$ we obtain $\Theta(\frac{N}{\log{N}}h(N)^{-1})$, and if $s>1$ this is  $\Theta(N h(N)^{-s})$. In particular, the highest mana node, i.e.,~$h(N)=1$, is queried $\Theta(N^s), \Theta(\frac{N}{\log{N}})$, or $\Theta(N)$ times, and might eventually be overrun by queries. Nodes whose rank is $\Theta(N)$ have to answer only $\Theta(1)$ queries. This is in contrast to the case $s=0$ where every node has the same mana and every node is queried in average a constant number of times.

The high mana nodes are therefore incentivized to gossip their opinions and not to answer each query separately. Since not all nodes can gossip their opinions (in this case every node would have to send  $\Omega(N)$ messages) we have to find a threshold when nodes gossip their opinions or not. 
If we assume that high mana nodes have higher throughput than lower mana nodes a reasonable threshold is $\log(N)$, i.e.,~only the $\Theta(\log(N))$ highest mana nodes do gossip their opinions, leading to $\Theta(\log{N})$ messages for each node in the gossip layer. In this case the expected number of queries the highest mana node, that is not allowed to gossip its opinions,  receives is  $\Theta((\frac{N}{\log{N}})^{s})$
if $s< 1$, $\Theta(\frac{N}{(\log{N})^2})$ if $s=1$, and $\Theta(\frac{N} {(\log{N})^{s}})$ if $s>1$.
In this case, nodes of rank between $\Theta(\log{N})$ and $\Theta(N)$ are the critical nodes with respect to message complexity.  

Another natural possibility would be to choose the threshold such that every node has to send the same amount of messages. In other words, the maximal number of queries a node has to answer should equal the number of messages that are gossiped. For $s<1$ this leads to the following equation
\begin{equation}
N^s h(N)^{-s} = h(N)    
\end{equation}
and hence we obtain that a threshold of order $N^{\frac{s}{s+1}}$ leads to $\Theta(N^{\frac{s}{s+1}})$ messages for every node to send. For $s>1$ one obtains similarly a threshold of $N^{\frac1{1+s}}$ leading to $\Theta(N^{\frac1{1+s}})$ messages. In the worst case, i.e.,~$s=1$, the message complexity for each node in the network is $O(\sqrt{N})$. 

We want to close this section with the remark that, as mentioned in Section \ref{sec:Zipf},  Zipf's law does mostly not apply on the entire range of the observations, but only on the upper tail regions of the observations. Adjustments of the above threshold and more precise message complexity calculations have to be performed in consideration of the real-world situation of the mana distribution. Moreover, the optimal choice of this threshold has also to depend on the structure of the network, and  the performances of the different nodes.

\section{Threat model}\label{sec:threat}

We consider the "worst-case'' scenario where adversarial nodes can exchange information freely between themselves and can agree on a common strategy. In fact, we assume that all Byzantine nodes are controlled by a single  adversary. We assume that such  an adversary holds a proportion $q$ of the mana and thus has a voting power $v_q=q$. 

\begin{figure}
\centering
    \includegraphics[width=0.8\textwidth,trim={0 0 0 0.cm},clip]{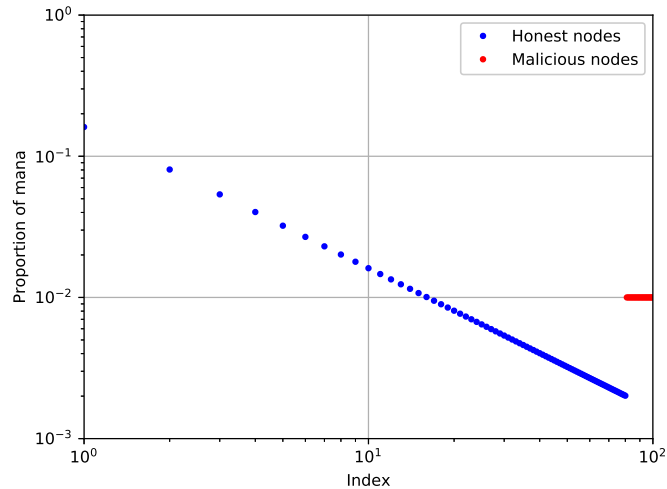}
  \caption{ Mana distribution with $s=1$, $N=100$ and $q=0.2$.}
  \label{fig:mana_distribution}
\end{figure}

In order to make results more comparable we assume that the adversary distributes the mana equally between its nodes such that each node holds $1/N$ of the total mana. Fig. \ref{fig:mana_distribution} shows an exemplary distribution of mana between all nodes. Nodes are indexed such that the malicious nodes have the highest indexes, while honest nodes are indexed by their mana rank.

We assume an "omniscient adversary", who is aware of all opinions and queries of the honest nodes. However, we assume that the adversary has no influence nor prior knowledge on the random threshold.

The adversary can take several approaches in influencing the opinions in the network. In a cautious strategy the adversary sends the same opinion to all enquiring nodes, while in a berserk strategy, different opinions can be sent to different nodes; we refer to  \cite{fpc,fpcsim} for more details. While the latter is more powerful it may also be easily detectable, e.g.,~see \cite{coordicide}. The adversary may also behave semi-cautious by not responding to individual nodes.

\subsection{Communication model}
We  have  to  make  assumptions  on  the  communication  model  of  the  FPC. We  assume  the  communication  between  two  nodes  to satisfy authentication, i.e.,~senders  and  receivers  are  who  they  claim  to  be,  and data  integrity,  i.e.,~data  is  not  changed  from  source  to  destination.  Nodes can also send a message on a gossip layer; these messages are then available to all participating nodes. All messages are signed by a private key of the sending node. 

As  we  consider  omniscient adversaries  we  do  not  assume confidentiality.   For  the  communication  of  the opinions between nodes we assume a  synchronous model. However, we want to stress that similar performances are obtained in a probabilistic synchronous model, in which for every $\varepsilon>0$ and $\delta>0.5$, a majority proportion $\delta$ of the messages is delivered within a bounded (and known) time, that depends on $\varepsilon$ and $\delta$, with probability of at least $1-\varepsilon$.  Due to its random nature, FPC still shows good performances in situations where not all queries are answered in due time. Moreover, the gossiping feature of high mana-nodes allows to detect  whether high mana nodes are eclipsed or are encountering communication problems.

\subsection{Failures}
In the case of heterogeneous mana distributions there are different possibilities to generalize the standard failures of consensus protocols: namely integration failure, agreement failure and termination failure. In this paper we consider only agreement failure since in the IOTA use case this failure turns out to be the most severe. In the strictest sense an agreement failure occurs if not all nodes decide on the same opinion. We will consider the $\alpha$-agreement failure; such a failure occurs if at least a proportion of  $\alpha$ nodes  differ in their final decision.


\subsection{Adversary strategies}

While \cite{fpc} studies robustness of FPC against all kinds of adversary strategies,  \cite{fpcsim} proposes several concrete strategies in order to perform numerical simulations. In particular,  \cite{fpcsim} introduced the  cautious \textit{inverse voting strategy} (IVS) and the berserk \textit{maximal variance strategy} (MVS). It was shown that, as analytically predicted in \cite{fpc}, the efficacy of the attacks is reduced when a random threshold is applied. The studies also show that the berserk attack is more severe, however in the presence of the random threshold the difference to IVS is not significant. Moreover, in Section \ref{sec:berserk} we propose efficient ways to detect berserk behavior.
The simpler dynamic of the IVS may also allow to approach the protocol more easily from an analytical viewpoint. For these reasons, we consider in this paper only a  cautious strategy that is an adaption of the IVS to the setting of mana.

\subsubsection*{manaIVS}
We consider the cautious strategy where the adversary transmits at time $t+1$ the opinion of the mana-weighted minority of the honest nodes of step $t$. More formally,
the adversary chooses
\begin{equation}
    \argmin_{j\in\{0,1\}} \sum_{i=1}^N m_i \mathbf{1}\{ s_i(t)=j  \}
\end{equation}
as its opinion at time $t+1$. We call this strategy the \textit{mana weighted inverse vote strategy} (\textbf{manaIVS}). 

\section{Improvements of FPC}\label{sec:improvements}
We suggest several improvements of the Vanilla FPC described in \cite{fpc}.


 
\subsubsection*{Fixed threshold for last rounds} In the original version of FPC nodes query at random including itself and finalize after having the same opinion for $\verb?l?$ consecutive rounds \cite{fpc}. We analyzed various situations when the Vanilla FPC encountered failures. One key finding was that the randomness of the threshold has sometimes a negative side effect. In fact, due to its random nature it will from time to time show abnormal behavior.\footnote{This is a common phenomenon for stochastic processes in random media; e.g.,~see \cite{dHo:00}.} In order to counteract this effect we can fix the threshold to a given value, e.g.,~$\tau=0.5$, for the last $\verb?l?_2$ rounds. The initial $\verb?l?-\verb?l?_2$ rounds enable the original task of FPC to create an honest super majority even in the presence of an adversary. Once a super majority is formed a simple majority rule is sufficient for the network to finalize on the same opinion, while the likelihood of nodes switching due to unusual behavior of the threshold is decreased significantly.

\subsubsection*{Bias towards own opinion}
In Section \ref{sec:fairness} we showed that with the introduction of mana as a Sybil protection we can adopt the FPC protocol in a fair manner by querying nodes with probability proportional to their mana. However,  this can lead to agreement failures if a mana high node over-queries the adversary in round $\verb?l?$. Part of the network would then finalize the opinion, while the mana-weighted majority of nodes could still switch their opinion. In an extreme situation it is possible that a node that holds the majority of the funds adjusts its opinion according to a minority of the funds, which is undesirable. 

In order to prevent this we propose the following adaption. Each node biases the received mean opinion $\eta$ to its current own opinion. More specifically, a node $j$ can calculate its $\eta$-value of the current round $i$  by 
$$\eta_{i}(t+1)= m_j s_{i}(t) + (1-m_j)\eta^*_{i}(t+1),$$
where $m_j$ is $j$'s proportion of mana and $\eta^*_{i}(t+1)$ is the mean opinion from querying nodes without self-query.

\subsubsection*{Fixed number of effective queries}
As discussed in Section \ref{sec:fairness} in order to facilitate a fair quorum (thereby preventing game-ability) we select for a given vote a node at random with a probability proportional to the mana. If a node is selected $m$ times it is given $m$ votes (of which all would have the same opinion). However this can lead to a quorum with a \textit{population} of nodes $k_{\text{diff}}<k$, in particular in scenarios where $N$ is low or $s$ is large. Furthermore, if there is a fixed bandwidth reserved to ensure the correct functioning of the voting layer, individual nodes could  regularly under-utilize this bandwidth since the communication overhead is proportional to $k_{\text{diff}}$. We can alleviate this deficit by increasing $k$ dynamically to keep $k_{\text{diff}}$ constant, and thereby improve the protocol by increasing the effective quorum size $k$ automatically.

Through this approach the protocol can adopt dynamically to a network with fewer nodes or different mana distributions.

\section{Berserk detection}\label{sec:berserk}

Since berserk strategies are the most severe attacks, e.g.,~\cite{fpc,fpcsim}, the security of the protocol can be improved if berserk nodes can be identified and removed from the network. We, therefore, propose in this section a mechanism that allows to detect berserk behavior. This mechanism is based on a "justification of opinion" where  nodes exchange information about the opinions received in the previous rounds. As the set of queried nodes changes from round to round this information does not necessarily allow a direct direction of a berserk behavior but berserk behavior is detectable indirectly with a certain probability. Upon discovering malicious behavior, nodes can gossip the proofs of this behavior, such that all  other honest nodes can ignore the berserk node afterwards.  

\subsection{The berserk detection protocol}
We allow that a node can ask a queried node for a list of opinions received during the previous round of FPC voting. We call such a list a vote list and write $v$-list. A node may request for it in several ways. For example, the full response message to the request of a $v$-list and the opinions could be comprised of the opinion in the current round and the received opinions from the previous round. We do not require nodes to apply this procedure for every member of the quorum or every round. For instance, each node could request the list with a certain probability or if it has the necessary bandwidth capacity available. Furthermore, we can set an upper bound on this probability on the protocol level so that spamming of requests for $v$-lists can be detected. We denote this probability that an arbitrary query request includes a request for a $v$-list by $p_B$.

A more formal understanding of the approach is the following: assume that in the last round a node $y$ received $k$ votes, submitted by nodes $z_1,...,z_k$. If a node $x$ asks $y$ for a $v$-list, then $y$ sends votes submitted by $z_1,...,z_k$ along with the identities of $z_1,...,z_k$ but without their signatures. This reduces the message size. Node $x$ compares the opinions in the $v$-list submitted by $y$ with other received $v$-lists. If $x$ detects a node that did send different opinions it will ask the corresponding nodes for the associated signatures in order to construct a proof of the malicious behaviour. Having collected the proof the honest node gossips the evidence to the network and the adversary node will be dropped by all honest nodes after they have verified the proof.

Note that a single evidence for berserk behaviour is sufficient and that further evidence does not yield any additional benefit. 

\subsection{Expected number of rounds before detection}

To test how reliable this detection method is and what the communication overhead would be, we carry out the following back-of-the-envelope  calculations for $s=0$ and $s>0$. We are interested in the probability of detecting a berserk adversary since the inverse of this probability equals the estimated number of rounds that are required to detect malicious behaviour of a given node. 

Let us start with $s=0$ and consider  the following scenario. Among $N$ nodes there is a single berserk node $B$. In the previous round, the adversarial node is (in expectation) queried $k$ times. To see this note that in the case of $s=0$, nodes are queried with uniform probability and every node has to receive on average the same number of queries.  Furthermore, the berserk node sends $f$ replies with opinion 0 to the group of nodes $G_0$ and $(k-f)$ replies with opinion 1 to the group of nodes $G_1$.

The probability that a node $x$ receives $v$-lists that allow for the detection of the berserk node is in this case bounded below by
\begin{align*}
&P(x {\text{ receives $v$-list from }} G_0 {\text{ and }} G_1) \cr &\geq  2  \binom{k}{2} p_B^2 \frac{f}{N}\cdot \frac{k-f}{N-1} \cdot  \frac{N-k}{N-2} \cdots \frac{N-2k+3}{N-k+1}=\gamma_0.
\end{align*}
The probability that some node detects the berserk behaviour satisfies
\begin{align*}
P({\text{some node detects malicious node}})\geq 1-(1-\gamma_0)^{N-1}.
\end{align*}

For example, in a system with $N=1000$, $k=20,  p_B=0.1$ and  $f=k-f =10$ the detection probability is bounded below by  $0.23$. Assuming that the full FPC voting (i.e.,~a voting cycle) for a conflict takes about $15$ rounds, berserk nodes can be detected within one FPC  voting cycle with high probability. 

Precise calculations are more difficult to obtain for $s>0$ and we give rough bounds instead.
Let us assume that $B$ holds the mana proportion $m_B$. In the case of mana, i.e.,~$s>0$, it is not the number of nodes, that are querying the berserk node, that is essential, but their mana. The probability that any given honest node queries the berserk node is at least $m_B$, which implies that the average sum of mana of honest nodes that query the berserk node is at least $m_Q=m_B(1-m_B)$. We assume that we can split up these nodes into two groups $G_0$ and $G_1$ of  equal mana weight, i.e.,~$m_{G1}=m_{G2}$. The berserk node answers $0$ to the nodes in $G_0$ and $1$ to the ones in $G_1$.  Then the probability that an honest node $x$ queries and requests a $v$-list from a node from the group $G_i$ ($i=0,1$) is at least $p_B m_Q/2$. Moreover,
\begin{align*}
&P(x {\text{ receives $v$-list from }} G_0 {\text{ and }} G_1) \cr &\geq 2 \left(\frac{p_B m_Q}2\right)^2=\gamma_1.
\end{align*}
Similarly to above,
\begin{align*}
P({\text{some node detects malicious node}})\geq 1-(1-\gamma_1)^{N-1}.
\end{align*}

For instance, if $N=1000, p_B=0.1$ and  $m_B=0.2$ the detection probability is  greater than $0.12$. Note that the above bound holds already for $k=2$. Hence, higher values of $k$ will lead to detection probabilities close to $1$.

\section{Heuristic for choosing the quorum size}\label{sec:k}

An important parameter that dominates the performance is the quorum size $k$. It may be chosen as large as the network capacity allows, in a dynamic fashion or as small as security allows to be sustainable. Previous results, e.g.~\cite{fpc} and \cite{cruise2013probabilistic}, show that an increase of $k$ decreases the failures rates exponentially. 
Let us give here some heuristic probabilistic bounds on what kind of values of $k$ may be reasonable. Here we consider only the Vanilla FPC but note that the same behaviour occurs for the changed protocol. The case $s=0$ can be treated analytically as follows. 

One disadvantage of the majority voting is that even if there is already a predominant opinion present in the network, e.g.,~opinion 1 if $p>\tau$, that a node picks by bad chance too many nodes of the minority opinion.

Let $p$ be the  average opinion in the network and $\tau$ the threshold with which a node decides whether to choose the opinion 1 or 0 for the next round. More specifically if more than  $\tau k$ nodes respond with 1 the node selects 1, or 0 otherwise. The number of received $1$ opinions follows a Binomial distribution $\mathcal{B}(k, p)$. 
Hence, the probability for a node to receive opinions that result in an $\eta$-value leading to the opinion 0 is given by 
$$
P_{0,k}(\tau)=P(Y\leq \lfloor \tau k\rfloor)=\sum_{m=0}^{\lfloor \tau k\rfloor} \binom{k}{m}p^m(1-p)^{k-m},
$$
where $Y\sim \mathcal{B}(k, p).$
As we are interested in the exponential decay of the latter probability as $k\to \infty$ we use a standard large deviation estimate, e.g.,~\cite{dHo:00}, to obtain for $\tau < p$:
\begin{equation}
    P_{0,k}(\tau)  \approx e^{-k I(\tau)},
\end{equation}
with rate function 
\begin{equation}
    I(\tau)= \tau \log\left(\frac{\tau}{p}\right) + (1-\tau)  \log\left(\frac{1-\tau}{1-p}\right).
\end{equation}
This shows an exponential decay of $P_{0,k}(\tau)$ in $k$ and that the rate of decay depends on the "distance" between $p$ and $\tau$.

\begin{figure}
\centering
    \includegraphics[width=0.8\textwidth,trim={0 0 0 0cm},clip]{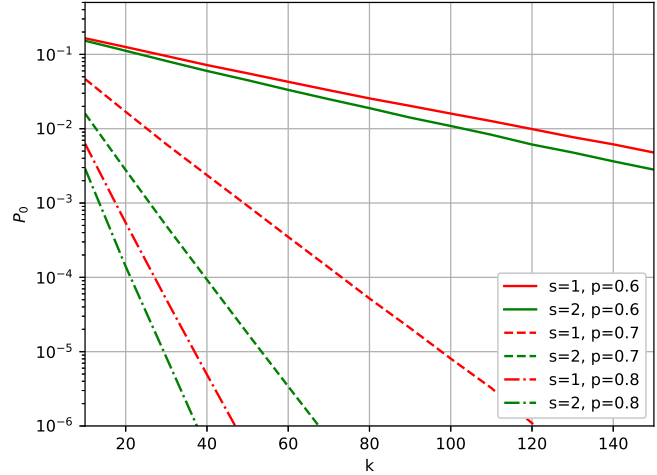}
    \caption{ Probability for a node to choose the opinion 0 for $\tau=0.5$ in the mana setting.}
    \label{fig:P_0mana}
\end{figure}

An exact calculation in the mana setting of these probabilities is more difficult to obtain. We consider the situation where the top mana holders have opinion $1$ and the remaining nodes have opinion $0$ such that a proportion $p$ of the mana has opinion $1$. Fig. \ref{fig:P_0mana} shows estimates, obtained by Monte-Carlo simulations, of the probability that the highest mana node will switch to opinion $0$.

\section{Simulation results}\label{sec:simresults}

We perform simulation studies with the parameters given in Table \ref{tab:defaultparams} and study the $1\%$-agreement failure.  In order to make the study of the protocol numerically feasible we choose the system parameters such that a high agreement failure is allowed to occur. However as we will show the parameters can be adopted such that a significantly lower failure rate can be achieved.

 The source code of the simulations is made open source and available online.\footnote{\url{https://github.com/IOTAledger/fpc-sim}}

\begin{figure}
\centering
\begin{tabular}{llc}
\hline
 & Parameter & Value\\ \hline\hline
$N$ & Number of nodes & 1000\\ \hline
$p_0$ & Initial average opinion & 0.66 \\\hline
$\tau$ & Threshold in first round & 0.66 \\\hline
$\beta$ & Lower random threshold bound& 0.3\\\hline
$k$& Quorum size & 20 \\\hline
$\verb?l?$  & Final consecutive round & 10 \\ \hline
$\verb?maxIt?$ & Max termination round & 50 \\ \hline
$q$ & Proportion of adversarial mana & 0.25 \\\hline
$\alpha$ & Minimum proportion of mana & 0.01 \\
& for agreement failure &\\\hline
\hline
\newline
\end{tabular}
\label{tab:defaultparams}
\caption{Default simulation parameters}
\end{figure}


The initial opinion is assigned as follows. The highest mana nodes that hold together more than $p_0$ of the mana are assigned opinion $1$ and the remaining opinion $0$. More formally, let 
\begin{equation*}
    J := \min\{j: \sum_{i=1}^j m_i > p_0\},
\end{equation*}
then $s_i(0)=1$ for all $i\leq J$ and $s_i(0)=0$ for $j> J$.

\begin{figure}
\centering
    \includegraphics[width=0.8\textwidth,trim={0 0 0 0cm},clip]{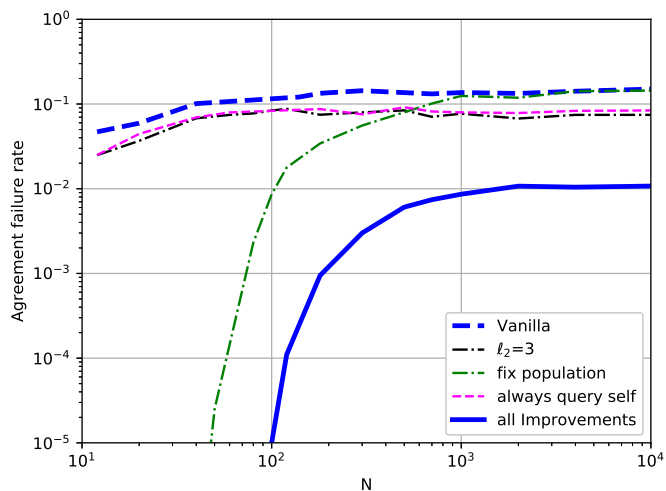}
  \caption{ Agreement failure rates with $N$, for $s=0$. The improvements from Section \ref{sec:improvements} are applied individually. }
  \label{fig:N}
\end{figure}

We investigate a network with a relatively small quorum size, $k=20$ and a homogeneous mana distribution ($s=0$). The adversary is assumed to hold a large proportion of the mana with $q=0.25$. Fig. \ref{fig:N} shows the agreement failure rate with $N$. We observe that the improvements from Section \ref{sec:improvements} increase the protocol significantly for the lower range of $N$. For a large value of $N$ the improvements are still of the order of one magnitude. 

\begin{figure}
\centering
    \includegraphics[width=.8\textwidth,trim={0 0 0 0cm},clip]{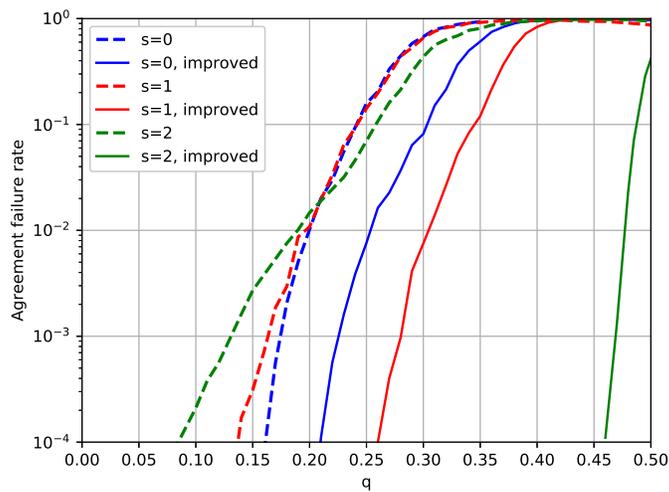}
  \caption{ Agreement failure rates with $q$ for three different mana distributions. \newline}
  \label{fig:q}
\end{figure}

Fig. \ref{fig:q} shows the agreement failure rate with the adversaries' mana proportion $q$. First, we can see that for the vanilla version the protocol performance remains approximately the same for small values of $s$, however for $s=2$ we can observe a deterioration in performance. This effect may be explained by the skewness of the Zipf law, leading to a more centralized situation where high mana nodes opinion are susceptible to sampling effects described in Section \ref{sec:improvements}.  

We can also observe that the improvements enable the protocol to withstand a higher amount $q$ of adversarial mana and that for most values of $q$ the improvement is at least one order of magnitude. As we increase $s$ we can observe an agreement failure that is several orders of magnitudes smaller than without the improvements.


\begin{figure}
\centering
    \includegraphics[width=0.8\textwidth,trim={0 0 0 0cm},clip]{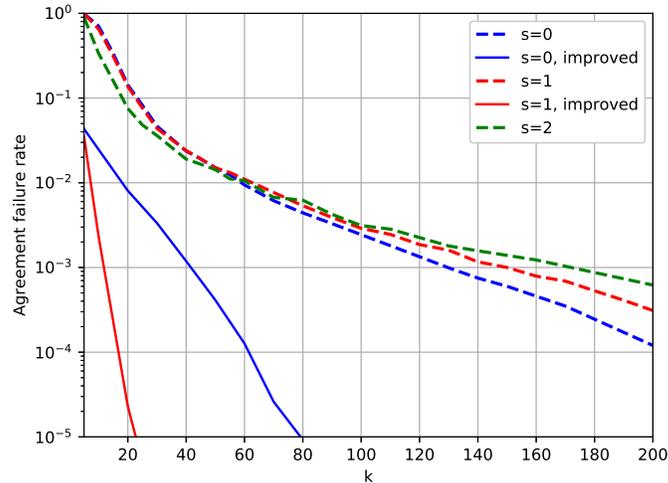}
  \caption{ Agreement failure rates with $k$. \newline}
  \label{fig:k}
\end{figure}

Fig. \ref{fig:k} shows the failure rate with the quorum size $k$. As discussed in Section \ref{sec:k} the probability for a node to select the minority opinion in a given round decreases exponentially with $k$ and this trend is also well reflected in the agreement failure rate, apart for small values of $k$. We show that the improvement of the failure rate becomes increasingly pronounced as the quorum size is raised. In Vanilla FPC the improvement decreases in the query size. Interesting to note that for small query sizes ($k\leq 60$), the centralized situation, $s>1$, is more stable against attacks, but for larger $k$ the centralized situations become more vulnerable than the less centralized ones. The improved FPC clearly performs better and the improvement of the agreement rate is more important as $s$ increases. 

Finally, for $s=2$ no failures are found in $10^6$ simulations for the improved algorithm, i.e.,~the failure rate is less than $10^{-6}$. This is in agreement with the  performance increase observed in Fig.~\ref{fig:q}.

We want to highlight that the experimental study above is only the first step towards a precise understanding of the protocol. There are not only many numerous parameters of the protocol itself, different ways to distribute the initial opinions, other types of failures to consider, but also many possible attack strategies that were not studied in this paper. We refer to \cite{fpcsim} for a more complete simulation study on the Vanilla FPC and like to promote research in the direction of \cite{fpcsim} for the FPC with weighted votes.

\section{Discussions}\label{sec:discussion}
A main assumption in the paper is that every node has a complete list of all other nodes. This assumption was made for the sake of simplicity. We want to stress out that in \cite{fpcsim} it was shown, for $s=0$, that in general it is sufficient that every node knows about $50\%$ of the other nodes. These results transfer to the setting $s>0$ in the sense that a node should know about nodes that hold at least $50\%$ of the mana. In many applications it is reasonable that all large mana nodes are publicly known and that this assumption is verified. 

Another simplification that we applied in the presentation of our results is that we assumed that the mana of every node is known and that every node has the same perception of mana. However, such a consensus on mana is not necessary. Generally, it is sufficient if different perceptions of  mana are sufficiently close. The influence of such differences on the consensus protocol clearly depends on the choice of parameter $s$ and may be controlled by adjusting the protocol parameters. However, a detailed study of the above effects is beyond the scope of the paper and should be pursued in future work.

For the implementation of FPC in the Coordicide version of IOTA, \cite{coordicide}, it is important to note that the protocol, due to its random nature, is likely to perform well even in situations where the Zipf law is partially or even completely violated.

The fairness results in Section  \ref{sec:fairness} concern the Vanilla FPC.  Similar calculations for the adapted versions are more difficult to obtain and beyond the scope of this paper. In particular, the sampling is no longer a sampling with replacement, but the sampling is repeated until $k$ different nodes are sampled; we refer to \cite{raj1958}  for a first treatment of the difference of these two sampling methods. The introduced bias towards its own opinion likely increases the voting power with respect to its own opinion but does not influence the voting power towards other nodes. Due to this fact and that linear weights are the most natural choice, we propose this voting scheme also for the adapted version.

\section*{Acknowledgment}
We are  grateful to all members of the \emph{coordicide team} for countless valuable discussions and comments on earlier versions of the manuscript.

\bibliographystyle{abbrv}
\bibliography{bibliography}

\end{document}